%% file: alpha_rich.tex
\documentclass[aps,prc,showpacs,preprint,superscriptaddress]{revtex4}
\usepackage{graphicx}
\usepackage{dcolumn}
\usepackage{amsmath}
\begin{document}

% Use the \preprint command to place your local institutional report
% number in the upper righthand corner of the title page in preprint mode.
% Multiple \preprint commands are allowed.
% Use the 'preprintnumbers' class option to override journal defaults
% to display numbers if necessary
%\preprint{}

%Title of paper
\title{Nuclear Reactions Important in Alpha-Rich Freezeouts}

% repeat the \author .. \affiliation  etc. as needed
% \email, \thanks, \homepage, \altaffiliation all apply to the current
% author. Explanatory text should go in the []'s, actual e-mail
% address or url should go in the {}'s for \email and \homepage.
% Please use the appropriate macro foreach each type of information

% \affiliation command applies to all authors since the last
% \affiliation command. The \affiliation command should follow the
% other information
% \affiliation can be followed by \email, \homepage, \thanks as well.
\author{George C. Jordan, IV, Sanjib S. Gupta, Bradley S. Meyer, Lih-Sin The}
\email[]{gjordan@clemson.edu}
\homepage[]{http://photon.phys.clemson.edu/nucleo}
%\thanks{}
%\altaffiliation{}
\affiliation{Department of Physics and Astronomy, Clemson University,
Clemson, SC 29634-0978}

%Collaboration name if desired (requires use of superscriptaddress
%option in \documentclass). \noaffiliation is required (may also be
%used with the \author command).
%\collaboration can be followed by \email, \homepage, \thanks as well.
%\collaboration{}
%\noaffiliation

\date{\today}

\begin{abstract}
\input{Abstract}
\end{abstract}

% insert suggested PACS numbers in braces on next line
% insert suggested keywords - APS authors don't need to do this
%\keywords{}

%\maketitle must follow title, authors, abstract, \pacs, and \keywords
\maketitle

% body of paper here - Use proper section commands
% References should be done using the \cite, \ref, and \label commands
%\section{}
% Put \label in argument of \section for cross-referencing
%\section{\label{}}

\input{Introduction}

\input{Calc}

\input{Web}

\input{Display}

\input{Explosion}

\input{Observables}

\input{56Cotext}

\input{57Cotext}

\input{59Nitext}

\input{55Fetext}

\input{Conclusion}

\begin{acknowledgments}
This research was supported by NASA grant NAG5-4703, NSF grant AST-9819877,
and a DOE SciDAC grant. The authors acknowledge helpful discussions with D. D. Clayton, M. D. Leising and P. Mohr.
\end{acknowledgments}

\bibliography{clemson}

\appendix*

\input{Perturbation}

\clearpage

\input{Isobar56}

\input{Isobar57}

\input{Isobar59}

\input{Isobar55}

\input{57Co}

\input{59Ni}

\input{55Fe}

\input{55Fe2}

\begin{figure}
\includegraphics{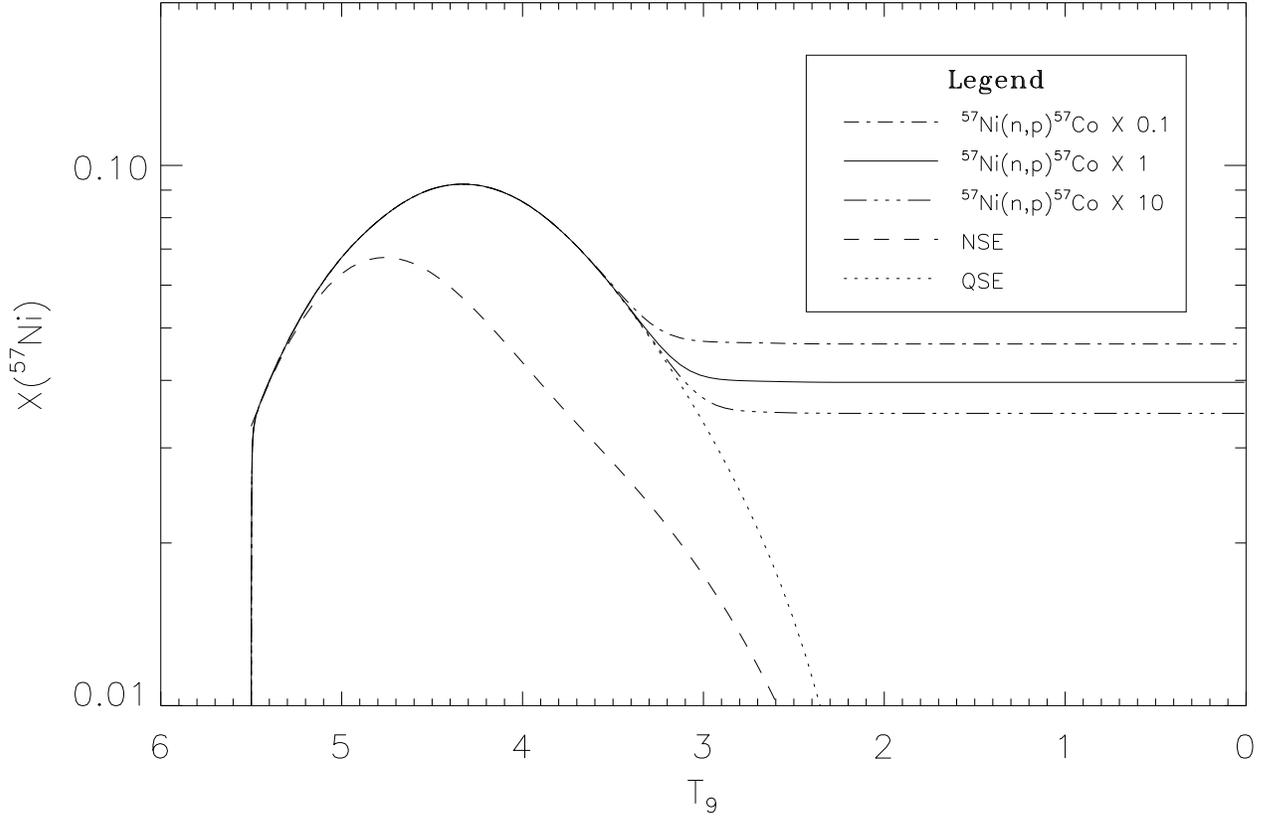}
\caption{\label{fig:57Co}Mass fraction of $^{57}$Ni versus T$_9$ for different values of the reaction rate $^{57}$Ni(n,p)$^{57}$Co and in NSE and QSE. The value of the reaction rate near T$_9 \approx$ 3.5 governs when $^{57}$Ni breaks out of QSE.}
\end{figure}

\begin{figure}
\includegraphics{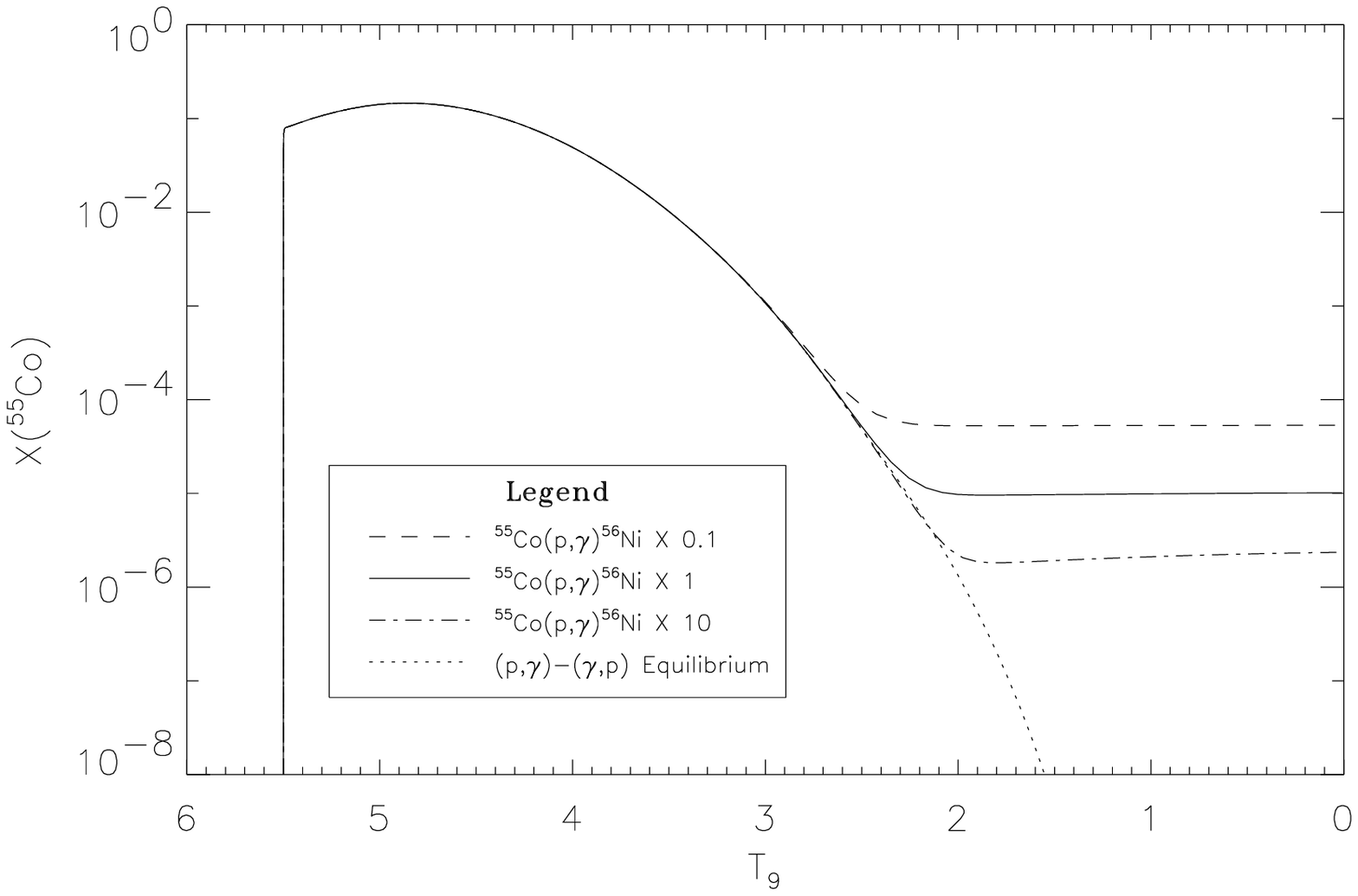}
\caption{\label{fig:55fe}Mass fraction of $^{55}$Co versus T$_9$ for different values of the reaction rate $^{55}$Co(p,$\gamma$)$^{56}$Ni and in (p,$\gamma$)-($\gamma$,p) equilibrium.  The value of the reaction rate near T$_9 \approx$ 2.5 determines when $^{55}$Co breaks out of (p,$\gamma$)-($\gamma$,p) equilibrium with $^{56}$Ni.}
\end{figure}

\end{document}

%% file: Abstract.tex
The alpha-rich freezeout from equilibrium occurs during the core-collapse explosion of a massive
star when the supernova shock wave passes through the Si-rich shell of the star.  The nuclei are heated to high
temperature and broken down into nucleons and $\alpha$ particles.   These subsequently reassemble as the material
expands and cools, thereby producing new heavy nuclei, including a number of important supernova observables. In
this paper we introduce two web-based applications. The first displays the results of a
reaction-rate sensitivity study of alpha-rich freezeout yields. The second allows the interested reader to run 
paramaterized explosive silicon burning calculations in which the user inputs his own parameters. These
tools are intended to aid in the identification of nuclear reaction rates important for experimental study.  We then
analyze several iron-group isotopes ($^{59}$Ni, $^{57}$Co, $^{56}$Co, and $^{55}$Fe) in terms of their roles as
observables and examine the reaction rates that are important in their production.

%% file: Introduction.tex
\section{Introduction\label{sec:intro}}
Progress in the science of stars and their nucleosynthetic processes
relies on the continued interplay of astronomical observations and astrophysical
modeling.  Observations of abundances of chemical species, elemental or isotopic,
constrain astrophysical models while the models, in turn, provide a framework for interpreting the observations.
It has long been clear, however, that uncertainities in the input physics to the models
limit them and their usefulness in interpreting abundance observations.  Nuclear reaction rates are key inputs into the
astrophysical models, and, though many are measured, most have not, and modelers must therefore rely on theoretical
predictions of the value of these rates.  Recent theoretical reaction-rate predictions have proven fairly accurate
(to within a factor of a few of the actual rate value where subsequently
measured--e. g., \cite{2000PhRvL..84.1651S}), nevertheless
experimental results are usually desirable and often crucial.  A third essential effort
in nuclear astrophysics is that of nuclear experimentalists who seek to provide better
input into astrophysical models by measuring the rates of nuclear reactions or nuclear properties that improve
theoretical estimates of the rates.  Because of the cost in time, effort, and
financial resources in performing the necessary experiments, however, the task of the nuclear experimentalists
is greatly aided if the astrophysical significance of a particular nuclear reaction can be clearly demonstrated.  This
in turn requires demonstration of the dependence of the predicted value of an astronomical observable in an
astrophysical model on the value of the reaction rate.  It is clear that observers, modelers, and
experimentalists should all play an important role in the planning of nuclear astrophysics experiments, in
particular by ensuring that any nuclear reaction rate proposed for experimental study satisfy the following four
requirements~\cite{1998ApJ...504..500T}:

\begin{enumerate}

\item An appropriate astrophysical model of a nucleosynthesis process must exist.
\item An observable from that process, usually an abundance result, is either known or measurable.
\item The dependency of the value of the observable on the value of the nuclear cross section is demonstrable.
\item An experimental strategy for measuring the reaction rate, or at least using experimental data to better calculate
the reaction rate, should be available.

\end{enumerate}

A long-term goal of our program is to aid the dialogue among the three parties (observers, modelers, and
experimentalists) in the nuclear astrophysics community by making it easier to identify those nuclear reactions most in
need of experimental study.  In the present work, we focus on reaction rates important in the alpha-rich freezeout in
core-collapse supernovae.  We explore in some detail astronomical observables from this process and the nuclear
reactions that govern their nucleosynthesis.  In addition, we present web-based tools that allow any interested
researcher to explore reaction-rate sensitivities in the alpha-rich freezeout and thereby make a case for experiments
on nuclear reactions important for other observables (yet to be identified) from this process.  We hope that this work
can serve as a template for future work (by ourselves or others) on nuclear reactions important in other nucleosynthesis
processes.

This paper begins with an introduction to the alpha-rich freezeout and a description of our freezeout calculations
and the attendant reaction-rate sensitivity studies in \S \ref{sec:calculations}.
In \S \ref{sec:web} we introduce two web sites accessible from the main page \url{http://photon.phys.clemson.edu/gjordan/nucleo/}. The first website 
displays data from the sensitivity calculation performed on our alpha-rich freezeout model in order
to identify nuclear reactions that are important in the production of nuclei identified
as astrophysical observables.  With a particular isotope in mind, one may then use the web site to view the effect of
different reaction rates on the yield of that isotope. The second website 
can be used to calculate the effects of a varied
reaction rate on nucleosynthesis yields under conditions that differ from the conditions used in our sensitivity survey.
Control over several of the parameters in the explosive model is given to the user so that the effect of a particular
reaction rate on the network yields can be explored over a wide variety of conditions, thereby allowing the user to
strengthen his case for measuring a particular reaction.
Finally, due to their significance as astrophysical observables,
several isotopes from the iron-group nuclei are examined \S \ref{sec:observables}.

%% file: Calc.tex
\section{The Calculations\label{sec:calculations}}
Many processes contribute to the production of new nuclei in core-collapse supernovae, but one of the most important for
astronomy is the alpha-rich freezeout from equilibrium.
During the core collapse of a dying massive star, a shock wave develops as matter falls supersonically onto the
collapsed stellar core.  The shock, aided by a push from neutrinos from the cooling nascent neutron star, expands out into,
heats, and expels the overlying stellar matter.  In the initial heating of the innermost regions of the ejecta,
post-shock temperatures are sufficiently high that nuclei are broken down into nucleons and $\alpha$ particles.
As the material subsequently expands and cools, the
nucleons and $\alpha$ particles reassemble to form heavy nuclei.  Because of the fast expansion of the matter, however,
not all alpha particles reassemble, and, as a result, the final abundances freeze out with a significant number
$\alpha$ particles remaining, hence the name alpha-rich freezeout.  A number of significant astronomical observables
are produced in this process including $^{44}$Ti, $^{56}$Co, and $^{57}$Co.

In order to explore the sensitivity of alpha-rich freezeout yields
to variations in reaction rates, we utilized the Clemson nucleosynthesis
code \cite{1995ApJ...449L..55M}, which we have updated to employ
the NACRE \cite{1999NuPhA.656....3A} and NON-SMOKER
\cite{2000ADNDT..75....1R} rate compilations.  The network used (see \cite{1998ApJ...504..500T}) includes 376 species from neutrons up to $Z=35$ (Bromine) and 2,125 reactions among them.

We began the survey with calculations using reaction rates from the compilations. Guided by detailed models of such astrophysical
settings (e. g., \cite{1996ApJ...460..408T}), we chose an initial temperature of $T_9 = T/10^9 {\rm K} = 5.5$ and
initial density $\rho_0 = 10^7$ g/cm$^{3}$ in all calculations.  The matter was taken to expand exponentially
so that the density evolved with time $t$ as $\rho(t) = \rho_0 \exp(-t/\tau_{ex})$,
where the density
e-folding timescale $\tau_{ex} = 446\ {\rm s}/\sqrt{\rho_0} = 0.141$ s. We assume the relation $\rho \propto
T^3$ (radiation-dominated expansion), and took three
possible values of the neutron excess $\eta$, viz., 0, 0.002, and 0.006.
Since we considered the appropriate model for alpha-rich freezeout to be
the passage of a shock wave through $^{28}$Si-dominated matter, the inital composition was $^{28}$Si with enough
$^{29}$Si to give the appropriate value of $\eta$.

Upon completion of these reference calculations, we explored the
significance of a particular reaction by multiplying the
reference value for its rate by a factor of ten, and repeating the calculation
with the same expansion parameters as in the reference calculations.  We emphasize that, with this procedure, we
uniformly increase the rate at all temperatures by a factor of ten. The corresponding reverse reaction rate is also increased by the same factor at all temperatures. This is required by detailed balance to ensure the material reaches the appropriate equilibrium at high enough temperature and density.  We repeated this procedure for all
2,125 reactions and all values of initial $\eta$'s. We also repeated
this sequence with a multiplicative factor 0.1 on each of the
reactions.  The total number of alpha-rich freezeout calculations
performed was (number of reactions) $\times$ (number of different
values of $\eta$) $\times$ (number of multiplicative factors) + (number of reference
calculations) $ = 2,125 \times 3 \times 2 + 3 = 12,753$.

The effect of modifying a particular reaction can be determined from the ratio of the modified yield of a
particular species with the reference yield. Clearly with 2,125 reactions and 376 species, there are many
combinations to consider.  Rather than present several large tables, we
have opted to construct interactive web-based applications to display the results.

%% file: Web.tex
\section{\label{sec:web}Internet Applications}
In this section we introduce two web-based applications. Since the web sites have complete online help files, detailed instructions are not given here. We will simply introduce the web sites and some of their relevant features, along with instructions on how to reproduce the tables appearing in the following sections.

%% file: Display.tex
\subsection{\label{subsec:disp}Network Sensitivity Data Display}

The Sensitivity Data Display website is designed to
examine the results of varying the reaction rates involving a selected isotope.  Upon loading the
page, the user selects from one of six data sets. The data sets are specified by $\eta$ and by whether the reaction rates are increased or decreased (by a factor of 10 in each case). Next,
the user can generate lists sorted by increasing or decreasing modified-to-standard ratios for a selected isotope. An
isotope (the chosen observable) is selected by entering its atomic number and mass number in the appropriate fields.
A $\beta$-decay option is also available which permits the user to evolve the abundances to arbitrarily later
times upon cessation of the alpha-rich freezeout process.  (The numerical technique used to construct the exact
solutions to the coupled differential equations governing the $\beta$-decay of the abundances is described in the
Appendix).

The data in Tables~\ref{tab:57co},~\ref{tab:59ni}, and ~\ref{tab:55fe} of \S \ref{sec:observables} were collected
using this web site. To reconstruct Table~\ref{tab:55fe}, for example, the user selects the data set corresponding
to $\eta$=0.006 and a multiplicative factor of 0.1. For $^{55}$Fe, 26 is entered as the $Z$
value and 55 for the $A$ value.  We target an observation of the $^{55}$Fe abundance made 2 years after the
explosion, and so the user selects the $\beta$-decay box and enters an evolution time of 2 years into the time field. 

In order to display only the reactions that change the $^{55}$Fe
yield by more than 20\%, the ratio cut-off parameter is set to 0.20.
The ``Descending List'' button is then clicked to generate the
ratio list for $^{55}$Fe.  The first table generated produced gives the mass fractions of all $A=55$ species in the
network at the end of the $\eta=0.006$ reference alpha-rich freezeout calculation together with their beta-decay or
electron-capture lifetimes.  Isotopes highlighted in black are stable while those in pink are unstable against $\beta^+$ or electron capture. Isotopes highlighted in blue are subject to $\beta^-$-decay while green denotes instability via
both channels. 

In the second table, the column on the far right of the output contains the modified-to-standard ratios after an
interval of two years. The reactions which, when their rate is decreased by a factor of ten, result
in a minimum increase in the yield of $^{55}$Fe of 20\% are entered in the
table along with the respective ratios. Similarly, those reactions which, when their rate is decreased by a factor of
ten, result in at least a 20\% decrease in the $^{55}$Fe yield were also entered in the table along with their ratios.
The process was then repeated with a rate factor of 10 and $\eta$ of 0.006. The data for Tables~\ref{tab:57co}i
and~\ref{tab:59ni} were created in a similar manner, but for the isotopes $^{57}$Co and $^{59}$Ni.

Other options on the web site include (a) a table of the isotopes used in network calculations,(b) a list of
reaction rates incluced in the calculations and (c) a tool for the user to plot or make a table of the standard
reaction rates used in the calculations.  The user can use the reaction number from the reaction list and then input
that reaction into the reaction-rate engine to obtain either a plot or table of the rate as a function of temperature.
(Note that the rate we present may include a stellar enhancement factor to correct for the possibility of excited
states in the target nucleus.  The user must account for
this factor when comparing his rate with the one we used.)  Other links on the page are
to various related topics in the help file and to the NACRE and NON-SMOKER rate
compilation web sites.

%% file: Explosion.tex
\subsection{\label{subsec:explosion}Explosive Nuclear Burning}
The Explosive Nuclear Burning web site
simulates parameterized explosive nuclear burning over the web using the same reaction network as described above.
Among the variable parameters are $T_9$ (initial temperature in $10^{9}$ K),
$\rho$ (initial density in g/cm$^3$), $\tau_{ex}$ (the expansion time scale of the explosion), and the
initial mass fraction of each species. The user can also alter any reaction rate in the network
by a multiplicative factor as was done in \S~\ref{sec:calculations}. 

The data output
from the web site are the final mass fractions from the calculation. If a reaction rate is modified, the web site
performs two calculations, one with the modified reaction rate, and one with the standard reaction rates for
comparison. The data output in this case are the final mass fractions from both calculations along with their
modified-to-standard ratios. The web site offers several options for further analysis, including a $\beta$- decay
option to evolve the final mass fractions in the output.

This web site was used to produce the data from Tables ~\ref{tab:iso56},~\ref{tab:57iso},~\ref{tab:iso59},
and ~\ref{tab:55iso}.  To reconstruct these tables, the user performs the following steps.  First, the user selects the
default parameters, which correspond to an alpha-rich freezeout of silicon-dominated matter with an $\eta$ of 0.006
(the parameters the sensitivity calculations in \S~\ref{sec:calculations}).  When the calculation is executed, the 
appropriate mass fractions are inserted into the
``t = 0 years'' column of the Table. The $\beta$-decay box is then selected from the ``Final Mass Fraction Data List Controls''
table in the left hand frame and a time of 2 years entered. The ``Create List'' button is then clicked.
The web site returns a table with the value of all of the final mass fractions after 2 years.
The final mass fractions should agree with those in Tables~\ref{tab:iso56},~\ref{tab:57iso},~\ref{tab:iso59},
and ~\ref{tab:55iso}.

The data for Table~\ref{tab:55fe2} was also generated with this web site. To reconstruct this table, the user
performs the following steps.  First, the web site is reloaded into the
browser. Then the ``Edit'' button in the ``Reaction Rates'' row of the ``Parameter Controls'' table in the left hand
frame is clicked. This brings up an interface that allows the user to select a reaction rate to modify based on the
type of reaction preferred. Since the reaction of interest in Table~\ref{tab:55fe2} is $^{55}$Co($p,\gamma)^{56}$Ni
(a two nuclear species to one nuclear species reaction),
the radio button at the bottom of the table entitled ``$i + j \to k$'' is clicked.  Starting at the left hand side
of this table, the consituents of the reaction are chosen field by field. When all of the fields in the table are
selected, the ``Save Changes'' button at top (or bottom) of the page is clicked. The user is then prompted to input a
reaction rate multiplier. A value of 100 is entered and the ``Save Changes'' button clicked.
 The web site then presents a summary of the changes and the calculation is begun by clicking the
``Run Nucleosynthesis Code'' button. The outputted data are then evolved for two years as above and the
appropriate modified-to-standard ratio is entered into Table~\ref{tab:55fe2}. The web site is then reloaded and
the same procedure is carried out except 0.01 is entered as the multiplicative factor on the
$^{55}$Co(p,$\gamma$)$^{56}$Ni reaction rate.   These steps should reproduce the data in Table \ref{tab:55fe2}.

This web site has many other useful tools that are not described above but are documented in the on-line help files.
As shown, this web site allows the user to examine
the effects of a reaction on nucleosynthesis yields under a wide variety of circumstances.  Our hope
is that this easily accessible data will provide additional insight into the effects that uncertainties
in reaction rates have on the yields of the alpha-rich freezeout and other nucleosynthesis processes.

%% file: Observables.tex
\section{Nuclear Reactions Governing the Synthesis of Iron-Group Observables\label{sec:observables}}
In this section, we present some
results of our calculations drawn from the web sites described in
\S~\ref{sec:web}.  For brevity we limit the discussion to the $\eta$ =
0.006 calculations. The purpose is to illustrate how one may identify
relevant observables and identify their governing reactions using the
web sites.

At least three types of isotopic observables are relevant for stellar nucleosynthesis:
\begin{enumerate}

\item The bulk yields
are important for understanding galactic chemical evolution and solar system abundances.

\item  Radioactive species
such as $^{26}$Al and $^{44}$Ti  can be observed from space telescopes, and consequently
provide important constraints on their production sites.

\item Isotopic abundances in presolar meteoritic grains carry important information 
about the nucleosynthesis environments in which their isotopes formed and the
astrophysical settings in which the dust grains condensed (e. g.,
\cite{2002LPI....33.2018D,2002ApJ...578L..83C}).
\end{enumerate}

In the subsections to follow, we present
several isotopes from the iron-group nuclei for which we used the web-based applications in \S ~\ref{sec:web}
to obtain data.  These four isotopes are directly or indirectly linked
to all three types of observables listed above. We have chosen to analyze $^{57}$Co and $^{56}$Co because of
their prominent role as $\gamma$-ray observables and $^{59}$Ni and $^{55}$Fe for their potential as future
X-ray observables.  In addition, the abundances of the daughter isotopes of all four of these species may eventually be
measured in presolar grains.  We do not analyze $^{44}$Ti, a key alpha-rich freezeout observable, because this has been done previously \cite{1998ApJ...504..500T}.

For each species, we list any reaction rate that
produces at least a 20\% increase or decrease in the alpha-rich freezeout yield due to  a factor of 10 change in the
reaction rate (except for $^{57}$Co as explained below). Since we primarily consider the $\gamma$-ray or X-ray
observations of these species, we chose a time of analysis 2 years after the freezeout.   By this time,
overlying supernova should have become transparent to these radiations. Often, several
years pass, if not thousands, before data from astronomical observables can be collected. It is, therefore, up to the 
user to make his own analyses using appropriately chosen times in the web tools described in \S \ref{sec:web}.
For each observable considered, we list the important reactions, and for two particularly interesting reactions
we explore in some detail the reason for their effect on the chosen observable.

%% file: 56Cotext.tex
\subsection{\label{subsec:56Cotext}$^{56}$Co}

This species is produced primarily as the radioactive parent $^{56}$Ni ($\tau_{1/2}$ = 6.075 days) which decays
through $^{56}$Co ($\tau_{1/2}$ = 77.2 days) to $^{56}$Fe (see Table~\ref{tab:iso56}).
The overlying supernova largely obscures $\gamma$-rays from the decay of $^{56}$Ni, but the 2.598 Mev $\gamma$-rays
(for example) from the $^{56}$Co can escape and be detected~\cite{1990ApJ...357..638L}, making the latter species
a valuable alpha-rich freezeout observable\footnote{It should be clear that the bulk yield of $^{56}$Fe from the
alpha-rich freezeout may also be considered an observable since approximately half of the $^{56}$Fe in the solar
system was produced by alpha-rich freezeouts in core-collapse supernovae (e. g., \cite{1995ApJS...98..617T}).
Accurate alpha-rich freezeout yield predictions are required for Galactic chemical evolution models to reproduced
the solar system's supply of $^{56}$Fe.}.
The $\gamma$-rays from $^{56}$Co decay
can be used to determine the total yield of the parent ($^{56}$Ni) as well as the opacities of the outer envelope of
the supernova (for a review of $\gamma$-ray observables see~\cite{1998PASP..110..637D}).

From the web sites one finds that the abundance of $^{56}$Co after two years is dependent on the yield of $^{56}$Ni
from the supernova. We find that the yield of $^{56}$Ni immediately after the $\eta$ = 0.006 alpha-rich freezeout
is quite insensitive to factor-of-ten changes in the reaction rates. The largest effect was due to changes in the
triple-$\alpha$ reaction which produced about a 6\% change in the yield. This  6\% change propagates as a  6\% change
in the yield of $^{56}$Co after the decay of $^{56}$Ni. These are not significant changes and we conclude that the
calculated yields of $^{56}$Ni (and thus the yields of $^{56}$Co) are quite robust against any reaction rate
uncertainties. This result is expected since the production of $^{56}$Ni is dominated by the equilibrium phase of the
expansion in which the abundances are set by binding energies and nuclear partition functions (not by individual
reaction rates).

%% file: 57Cotext.tex
\subsection{\label{subsec:57co}$^{57}$Co}
A second important isotope is $^{57}$Co. This isotope is similar to $^{56}$Co in that its stable daughter  $^{57}$Fe
owes a significant portion of its synthesis to alpha-rich freezeouts.  From Table~\ref{tab:57iso} we see that $^{57}$Co
is primarily produced as the parent isotope $^{57}$Ni.  Gamma-rays from the decay of $^{57}$Co are observable with space detectors,
and they also power the supernova light curve at somewhat later times than $^{56}$Co~\cite{1995ApJ...450..805L}.
As shown, factors that affect the yield of $^{57}$Ni thus have an effect on the yield of $^{57}$Co.

The web sites show that the yield of $^{57}$Ni immediately after the alpha-rich freezeout is (like $^{56}$Ni) largely
insensitive to the value of any particular reaction rate because both are primarily equilibrium products. The reaction
with the most significant effect (for $\eta$= 0.006) is $^{57}$Ni(n,p)$^{57}$Co.  We have chosen to list this
reaction even though it does not produce more than a 20\% change because of the importance of $^{57}$Co as an observable. 
From Table~\ref{tab:57co} we see that this reaction produces at most a 17.7\% change in the yield of $^{57}$Ni for a factor of ten change in the rate.
The primary reaction rate governing the yield of $^{57}$Co for $\eta$ = 0.006 is $^{57}$Ni(n,p)$^{57}$Co
and a factor of a few uncertainty in the reaction rate results in a several percent uncertainty in the observable.

The reason $^{57}$Ni(n,p)$^{57}$Co has an effect on the yield of $^{57}$Ni (and thus the yield of $^{57}$Co) is that
it governs when $^{57}$Ni falls out of quasi-statistical equilibrium (QSE--see, for example, \cite{1998ApJ...498..808M}).
As shown in figure~\ref{fig:57Co}, the $^{57}$Ni abundance diverges from nuclear statistical equilibrium (NSE) early but remains in line with QSE expectations until it freezes out
below T$_9 \approx$ 3.5. The QSE favors $^{57}$Ni early, but as the material cools below $T_9 \approx 4.3$, the QSE
abundances shift to higher-mass nuclei; thus, the longer $^{57}$Ni remains in QSE, the lower its final abundance.
$^{57}$Ni falls out of QSE when the $^{57}$Ni(n,p)$^{57}$Co reaction becomes too slow, which
means that increasing the rate for this reaction causes $^{57}$Ni to remain in QSE longer and to have a lower final
abundance. Decreasing the rate means $^{57}$Ni falls out of QSE earlier and retains more of its originally
high QSE abundance.

%% file: 59Nitext.tex
\subsection{\label{subsec:59ni}$^{59}$Ni}
Both $^{59}$Ni and $^{55}$Fe have been chosen because they hold promise as detectable radioactive X-ray sources
(for a prospectus see~\cite{2001ApJ...563..185L}). In the decay of these isotopes by electron capture, a K-shell
vacancy occurs which is often filled by an electron previously occupying a higher-level bound state. The energy loss
from this event is carried away as an X-ray which, if detected, would not only provide data complementary to $\gamma$-ray
observations, but also shed light on the production of radioactive isotopes that emit no $\gamma$-rays.

The observable $^{59}$Ni, because of its 75,000 y half-life, should
produce a significant diffuse emission from the
interstellar medium similar to that observed from interstellar $^{26}$Al.  It may
also be detectable in close, individual supernova
remnants~\cite{2002NewAR..46..529L}. From Table~\ref{tab:iso59} we see
that the majority of $^{59}$Ni is produced in the alpha-rich freezeout
as $^{59}$Cu; thus, factors that affect the production of $^{59}$Cu
ultimately affect the production of the observable $^{59}$Ni. From the
web sites it is evident that there are several reactions which, when
changed by an order of magnitude, have a moderate effect on the
production of $^{59}$Cu (see Table~\ref{tab:59ni}).  The most
influential of these is the $^{59}$Cu(p,$\gamma$)$^{60}$Zn reaction. A
factor of 10 decrease in this reaction rate produces about a 50\%
increase in the yield, while a factor of 10 increase in the reaction
rate drops the yield of $^{59}$Cu by about 30\%. These are modestly significant
changes and show that $^{59}$Ni is sensitive to specific
reactions in the alpha-rich freezeout.

%% file: 55Fetext.tex
\subsection{\label{subsec:55fe}$^{55}$Fe}
Unlike $^{59}$Ni, $^{55}$Fe has a relatively short half life ($\tau_{1/2}$ = 2.73 yrs), making it only detectable in
very young supernovae (supernova 1987A remains a good target for detecting $^{55}$Fe~\cite{2001ApJ...563..185L}).
Since the detection of $^{55}$Fe could become available at very early times after the explosion (assuming the ejecta
is transparent to X-rays), its detection should give information on small-scale structure formation
in the ejecta. Further, it may help elucidate the velocity structure of the core and the velocity structure of $^{55}$Fe
~\cite{2002NewAR..46..529L}.

As can be seen from Table~\ref{tab:55iso}, $^{55}$Fe is predominantly produced as $^{55}$Co. The production of $^{55}$Co
in the alpha-rich freezeout with $\eta$ = 0.006 is most sensitive to the triple-$\alpha$ reaction and the
$^{55}$Co(p,$\gamma$)$^{56}$Ni reaction (Table~\ref{tab:55fe}). The triple-$\alpha$ reaction has about
a factor of two effect on the yield when a factor of 10 change in the triple-$\alpha$ rate is made.
The largest effect on the yield is seen in the dependence of the $^{55}$Co abundance on the
$^{55}$Co(p,$\gamma$)$^{56}$Ni reaction.
A factor of 10 decrease in this reaction rate increases the yield of $^{55}$Co by 526\% while an increase in the
reaction rate reduces the yield by almost 80\%.

The reason for the sensitivity of $^{55}$Co to the $^{55}$Co(p,$\gamma$)$^{56}$Ni reaction rate is similar to that of
the sensitivity of $^{57}$Ni to the $^{57}$Ni(n,p)$^{57}$Co reaction rate, but the effect is much larger.  The $^{55}$Co
remains in (p,$\gamma$)-($\gamma$,p) equilibrium with abundant $^{56}$Ni until $T_9$ drops below 3.0. As the temperature drops, the equilibrium shifts abundance from $^{55}$Co to
higher-mass nuclei (particularly $^{56}$Ni). Increasing the $^{55}$Co(p,$\gamma$)$^{56}$Ni reaction rate causes
this abundance shift to persist longer, which in turn leads to a lower final
$^{55}$Co yield. A lower value for the reaction rate leads to an earlier freezeout from QSE and a higher
final $^{55}$Co yield. From this we see that $^{55}$Co, and thus $^{55}$Fe (the relevant observable), are very
sensitive to the $^{55}{\rm Co}(p,\gamma)^{56}$Ni reaction rate.

%% file: Conclusion.tex
\section{Conclusion\label{sec:conclusion}}
The above described web sites are intended to be a starting point for identifying those nuclear reactions that
govern the production of astrophysical observables from the alpha-rich freezeout.  It is our hope that these
web sites will help motivate future measurements of these key reactions, especially as new observables from the
alpha-rich freezeout are identified.

While we identified and analyzed four key observables from the alpha-rich
freezeout, we expect new ideas or emphases in astrophysics to make new observables
available.  For example, as models of Galactic chemical evolution become increasingly refined, better yields from
massive stars will become necessary.  This means that any isotope that owes a significant portion of its solar system
abundance to the alpha-rich freezeout, such as $^{40}$Ca, $^{48}$Ti, $^{52,53}$Cr and the ones discussed above,
will become an observable.  Similarly, new technologies may also open new possibilities for observing isotopes.
As a possible example, $^{44}$Ca excesses found in presolar supernova silicon carbide (SiC)
X grains arose from condensation of live $^{44}$Ti \cite{1996ApJ...462L..31N}.  One
speculation is that much of this excess radiogenic $^{44}$Ca may be concentrated in small
titanium carbide (TiC) subgrains within the larger SiC X grains.  Such TiC subgrains
are known to exist in mainstream SiC grains that condensed in outflows from low-mass stars
\cite{1992LPICo.781....1B,2000M&PS...35.1157H};
however, in two SiC X grains studied so far, there is no
direct evidence for the presence of such TiC subgrains\cite{2002LPI....33.1297B}, though the
search continues.  If such subgrains do exist, they are likely to be dominantly comprised
of alpha-rich freezeout material.  If the new generation of secondary-ion mass spectrometers
are able to measure the isotopic abundances in the subgrains, then the alpha-rich freezeout
abundances of all titanium and carbon isotopes, as well as the isotopes of any element that
might condense in TiC, would become observables.  Those new observables will open up new governing reactions requiring
experimental study.

We view the web sites presented here as a first step for similar, future work on other nucleosynthesis processes.  The
goal in those future efforts, as in the present work, will be to help identify nuclear reactions that govern the
nucleosynthesis of astronomical observables.  As always, however, the challenge remains not only to identify
nuclear reactions that govern the synthesis of those observables, but also to identify the key astronomical
observables themselves.  Only by maintaining a healthy dialogue
among astronomers, astrophysical modelers, and nuclear experimentalists can we face that challenge and advance the
science of nuclear astrophysics.

%% file: Perturbation.tex
%\documentclass{article}
%\usepackage{amsmath}

%%%%%%%%%%%%%%%%%%%%%%%%%%%%%%%%%%%%%%%%%%%%%%%%%%%%%%%%%%%%%%%%%%%%%%%%%%%%%%%%%%%%%%%%%%%%%%%%%%%%
%TCIDATA{OutputFilter=LATEX.DLL}
%TCIDATA{Version=4.00.0.2312}
%TCIDATA{Created=Sunday, October 13, 2002 23:56:34}
%TCIDATA{LastRevised=Monday, October 14, 2002 01:40:00}
%TCIDATA{<META NAME="GraphicsSave" CONTENT="32">}
%TCIDATA{<META NAME="DocumentShell" CONTENT="Standard LaTeX\Blank - Standard LaTeX Article">}
%TCIDATA{CSTFile=40 LaTeX article.cst}
%\begin{document}

\section{\label{sec:perturbation}Analysis of the $\beta$-Decay Algorithm}

We now draw attention to the algorithm used by the Sensitivity
Data Display and the Explosive Nucleosynthesis web-sites to compute alpha-rich freezeout yields at a
specified point in time. Traditional approaches rely on Runge-Kutta or
implicit Euler techniques to integrate the coupled differential equations
(preferably with adaptive timestep adjustments). However, we have sought
a rapid, but exact solution.  To do this, we exploit the
elementary matrix solution to the system of linear coupled DEs that utilizes
eigenvalues and eigenvectors. For any matrix equation of the form 
\begin{eqnarray}
\overset{\bullet }{\mathbf{x}} &=&\mathbf{Ax} \\
\mathbf{x(}0) &\mathbf{=}&\mathbf{b}
\end{eqnarray}

the solution at all later times is given by:

\begin{equation}
\mathbf{x(}t)\mathbf{=}\sum c_{k}(\mathbf{\Phi }_{k}e^{{\omega}_k t}),
\end{equation}

where $\mathbf{\Phi }_{k}$ are the eigenvectors and $\omega _{k}$ the
corresponding eigenvalues of the matrix $\ \mathbf{A}$ (constructed from  $%
\beta -$decay rates and the application of mass conservation) which remain
constant during the evolution. The coefficients $c_{k}$ of the linear
combination are the elements of the product $\mathbf{T}^{-1}\mathbf{b}$
(where $\mathbf{T}$ is the matrix of eigenvectors of $\mathbf{A}$ ordered
strictly according to the index $``k"$) 

The technical challenge here is that $\mathbf{A}$ is singular and thus
reduction to upper Hessenberg form (the practical method of computing the
spectrum of a large asymmetric system) using QR Factorization fails. We may
circumvent this difficulty by employing very small perturbations to the
rates and thus disturbing the rate matrix away from singularity, as is the
norm for practical Numerical Linear Algebra applications in most fields of
engineering. However, we must be cognizant of the fact that a theorem of Watkins~\cite{1995...Watkins}
 requires the eigenvalues to be reasonably separated for the
system to be insensitive to perturbations. Unfortunately for the full 376$%
\times $376 matrix, the observed degeneracies are quite severe. 

On closer inspection, however, it is realized that the cause of the
degeneracies is quite simple: the matrix is composed of several
non-communicating blocks. Since only $\beta -$decays are considered, the
vector space decomposes into the \textit{direct sum} of these blocks, each
corresponding to a particular mass number. Blocks comprised of isotopes from
different mass numbers cannot traffic with each other since we exclude other
kinds of nuclear reactions. Several of these component submatrices are very
similar in their spectra, and therefore the full space suffers from many
closely spaced eigenvalues. 

The solution, however, is now obvious--we merely compute the spectrum of
each block, confident in the knowledge that the isotopes within a
particular block are not affected by those in any other. The complete state
vector of the abundances at any given time is then a simple concatenation of
the state vectors from the subspaces. Watkins' theorem is now
observed to hold for each block (though the blocks themselves are
singular as the full matrix was before) and abundances are output within seconds.

Obviously, the next step is  to extend the perturbation technique to
matrices which \textit{cannot} be decomposed into non-communicating
subspaces (blocks). Weakly coupled blocks may lend themselves to similar
perturbation analyses. Unfortunately Watkins' theorem does not
predict error bounds and the most we can do at this stage is carry out some
rough numerical experiments to check the sensitivity of the system to small
random perturbations. In our trials we found the effects to be negligible. Propagation of the perturbations caused fluctuations on the order of $10^{-15}$ in the most ill-conditioned blocks. Abundances of order $10^{-15}$ or less are quite small and for any practical purpose such small abundances are nonexistant. We conclude that the system was
very robust to the small kicks we gave it. 

We hope to incorporate the latest techniques in spectral computations on
large asymmetric systems as they become available, since nucleosynthesis
would be an exciting test-bed for these tools from the frontlines of
research in Numerical Linear Algebra. In the interim, perturbation
allows us to practically carry out an important calculation in seconds which
would otherwise have been impossible if we were to insist on a spectral solution. 

%\end{document}

%% file: Isobar56.tex
\begin{table}
\caption{\label{tab:iso56}Mass Fractions for A=56 with $\eta$ = 0.006}
\begin{ruledtabular}
\begin{tabular}{cdd}
	Isotope 
	&\multicolumn{1}{c}{t = 0 years}
	&\multicolumn{1}{c}{t = 2 years}
\\\hline
	 $^{56}$Ni
	&0.767
	&0 
\\
	  $^{56}$Co
	&3.356 \times 10^{-6}
	&1.190 \times 10^{-3}
\\
	 $^{56}$Fe
	&5.161 \times 10^{-12}
	&0.766
\\
\end{tabular}
\end{ruledtabular}
\end{table}

%% file: Isobar57.tex
\begin{table}
\caption{\label{tab:57iso}Mass Fractions for A=57 with $\eta$ = 0.006}
\begin{ruledtabular}
\begin{tabular}{cdd}
	
	Isotope 
	&\multicolumn{1}{c}{t = 0 years}
	&\multicolumn{1}{c}{t = 2 years}
\\\hline
	 $^{57}$Cu
	&0
	&0
\\
	  $^{57}$Ni
	&3.967 \times 10^{-2}
	&0
\\
	 $^{57}$Co
	&6.735 \times 10^{-7}
	&6.197\times 10^{-3}
\\
	$^{57}$Fe
	&3.429 \times 10^{-14}
	&3.348 \times 10^{-2}
\\
\end{tabular}
\end{ruledtabular}
\end{table}

%% file: Isobar59.tex
\begin{table}
\caption{\label{tab:iso59}Mass Fractions for A=59 with $\eta$ = 0.006}
\begin{ruledtabular}
\begin{tabular}{cdd}
	
	Isotope 
	&\multicolumn{1}{c}{t = 0 years}
	&\multicolumn{1}{c}{t = 2 years}
\\\hline
	 $^{59}$Ga
	&0	
	&0
\\
	  $^{59}$Zn
	&6.260 \times 10^{-14}
	&0
\\
	 $^{59}$Cu
	&1.887 \times 10^{-3}
	&0
\\
	$^{59}$Ni
	&4.065 \times 10^{-5}
	&1.928 \times 10^{-3}
\\
	$^{59}$Co
        &3.682 \times 10^{-13}
        &3.562 \times 10^{-8}
\\
\end{tabular}
\end{ruledtabular}
\end{table}

%% file: Isobar55.tex
\begin{table}
\caption{\label{tab:55iso}Mass Fractions for A=55 with $\eta$ = 0.006}
\begin{ruledtabular}
\begin{tabular}{cdd}
	
	Isotope 
	&\multicolumn{1}{c}{t = 0 years}
	&\multicolumn{1}{c}{t = 2 years}
\\\hline
	 $^{55}$Ni
	&3.306 \times 10^{-10}
	&0
\\
	  $^{55}$Co
	&1.013 \times 10^{-5}
	&0
\\
	 $^{55}$Fe
	&3.393 \times 10^{-10}
	&6.058\times 10^{-6}
\\
	$^{55}$Mn
	&0
	&4.072 \times 10^{-6}
\\
\end{tabular}
\end{ruledtabular}
\end{table}

%% file: 57Co.tex
\begin{table}
\caption{\label{tab:57co}Data from the $\eta$ = 0.006 network survey for $^{57}$Co}
\begin{ruledtabular}
\begin{tabular}{cdd}
	 Reaction
	& \multicolumn{1}{c}{Reaction Rate $\times$ 0.1}
	&  \multicolumn{1}{c}{Reaction Rate $\times$ 10.0}
	
\\\hline
	$^{57}$Ni(n,p)$^{57}$Co	
	&1.177
	&0.876
\\
\end{tabular}
\end{ruledtabular}
\end{table}

%% file: 59Ni.tex
\begin{table}
\caption{\label{tab:59ni}Data from the $\eta$ = 0.006 network survey for $^{59}$Ni}
\begin{ruledtabular}
\begin{tabular}{cdd}
	 Reaction
	& \multicolumn{1}{c}{Reaction Rate $\times$ 0.1}
	&  \multicolumn{1}{c}{Reaction Rate $\times$ 10.0}
	
\\\hline
	$^{59}$Cu(p,$\gamma$)$^{60}$Zn 
	&1.556
	&0.615
\\
	$2\alpha$($\alpha$)$^{12}$C
	&1.324
	&0.731
\\
	$^{59}$Cu(p,$\alpha$)$^{56}$Ni 
	&0.848
	&1.444
\\
\end{tabular}
\end{ruledtabular}
\end{table}

%% file: 55Fe.tex
\begin{table}
\caption{\label{tab:55fe} Data from the $\eta$ = 0.006 network survey for $^{55}$Fe}
\begin{ruledtabular}
\begin{tabular}{cdd}
	 Reaction
	& \multicolumn{1}{c}{Reaction Rate $\times$ 0.1}
	&  \multicolumn{1}{c}{Reaction Rate $\times$ 10.0}
	
\\\hline
	 $^{55}$Co(p,$\gamma$) $^{56}$Ni
	&5.258
	&0.232
\\
	 $^{59}$Cu(p,$\alpha$)$^{56}$Ni
	&1.170
	&0.738
\\
	 $^{59}$Cu(p,$\gamma$)$^{60}$Zn
	&0.690
	&1.133
\\
	 $2\alpha$($\alpha$)$^{12}$C
	&0.561
	&1.887
\\
\end{tabular}
\end{ruledtabular}
\end{table}

%% file: 55Fe2.tex
\begin{table}
\caption{\label{tab:55fe2}Data from web based Explosive Silicon Burning with $\eta$ = 0.006 for $^{55}$Fe}
\begin{ruledtabular}
\begin{tabular}{cdd}
	 Reaction
	& \multicolumn{1}{c}{Reaction Rate $\times$ 0.01}
	&  \multicolumn{1}{c}{Reaction Rate $\times$ 100.0}
	
\\\hline
	$^{55}$Co(p,$\gamma$)$^{56}$Ni	
	&34.544
	&5.469 \times 10^{-2}
\\
\end{tabular}
\end{ruledtabular}
\end{table}